\documentclass[prd,reprint,showpacs,showkeys]{revtex4-1}
\usepackage{amsfonts}
\usepackage{mdframed}
\usepackage{amssymb}
\usepackage{amsmath}
\usepackage{graphicx}
\usepackage[font={footnotesize,it}]{caption}

\begin{document}

\title{3+1-dimensional thin-shell wormhole with deformed throat can be
supported by normal matter}
\author{S. Habib Mazharimousavi}
\email{habib.mazhari@emu.edu.tr}
\author{M. Halilsoy}
\email{mustafa.halilsoy@emu.edu.tr}
\affiliation{Department of Physics, Eastern Mediterranean University, Gazima\u{g}usa,
Turkey. }
\date{\today }

\begin{abstract}
From physics standpoint exotic matter problem is a major difficulty in
thin-shell wormholes (TSWs) with spherical / cylindrical throat topologies.
We aim to circumvent this handicap by considering angular dependent throats
in $3+1-$dimensions. By considering the throat of the TSW to be deformed
spherical, i.e., a function of $\theta $ and $\varphi $, we present general
conditions which are to be satisfied by the shape of the throat in order to
have the wormhole supported by matter with positive density in the static
reference frame. We provide particular solutions / examples to the
constraint conditions.
\end{abstract}

\maketitle

\section{Introduction}

The seminal works on traversable wormholes and thin-shell wormholes (TSWs),
respectively by Morris and Thorne \cite{MT} and Visser \cite{V} both
employed spherical / cylindrical \cite{Rotating} geometry at the throats.
Besides instability of TSWs \cite{Stability} and the wormholes supported by
ghost scalar field \cite{Ghost} one major problem in this venture is the
violation of the null energy condition (NEC). Precisely, TSWs allow presence
of exotic matter at the throat with $\sigma <0$ in which $\sigma $ is the
energy density on the hypersurface of the throat. Several attempts have been
made to introduce the TSWs supported by normal matter of the kind with $%
\sigma >0$ in the framework of the Gauss-Bonnet theory of gravity \cite{GB}.
We aim in this study to seek for $\sigma >0$ against violation of NEC by
changing the spherical / circular geometry to more general, angular
dependent throats in the wormholes. That is, since NEC is violated a
different choice of frame may account a negative energy density. Motivation
for such a study originates from the consideration of Zipoy-Voorhees (ZV)
metrics which surpasses spherical symmetry with a quadrupole moment by
employing a distortion parameter \cite{ZV}. In brief, this amounts to
compress a sphere into an ellipsoidal form through a distortion mechanism.
This minor change contributes to the total energy and makes it positive in
the static reference frame under certain conditions. In local angular
intervals we confront still with negative energies in part but the integral
of the total energy happens to be positive. We recall that any rotating
system with spherical symmetry becomes axial in which by employing a similar
refinement of the throat we may construct wormholes with a positive total
energy. It is our belief that by this method of suitable choice of geometry
at the throat and in a special frame we can measure a positive energy.
Recently we have shown \cite{MH} that the flare-out conditions \cite{HV}
which were thought to be unquestionable can be reformulated. We must admit,
however, that although geometry change has positive effects on the energy
content this doesn't guarantee that the resulting wormhole becomes stable.
For the particular case of counter-rotational effects in $2+1-$dimensional
TSW we have shown that stability conditions are slightly improved \cite{MH2}%
. That is, when the throat consists of counter-rotating rings in $2+1-$%
dimensions the stability of the resulting TSW becomes stronger. This result
has not been confirmed in $3+1-$dimensional TSWs yet. Arbitrary angular
dependent throat geometries have also been considered by the same token
recently in $2+1-$dimensions \cite{MH3}. Therein a large class of wormholes
with non-circular throat shapes are pointed out in which positive energy
supports the wormhole. In the same reference we explain also the
distinctions (if any) by employing the ordinary time instead of the proper
time. Extension of this result to the more realistic $3+1-$dimensions makes
the aim of the present study. Numerical computation of our chosen ansatzes
yield positive total energy, as promised from the outset.

We start with the $3+1-$dimensional flat, spherically symmetric line element
in which a curved hypersurface is induced to act as our throat's geometry.
Such a hypersurface, $\Sigma \left( t,r,\theta ,\varphi \right) =0$, has an
induced metric satisfying the Einstein equations at the junction with the
proper conditions, obeying the flare-out conditions. No doubt, such an
ansatz is too general, for this reason they are restricted subsequently.
Static case, for instance eliminates time dependence in $\Sigma \left(
t,r,\theta ,\varphi \right) =0$. We derive the general conditions for such
throats and present particular anzatses depending on $\theta $ and $\varphi $
angles alone that satisfy our constraint conditions.

Organization of the paper goes as follows. In Section II we present in brief
the formalism for TSWs. Static TSWs follow in Section III where angular
dependent constraint conditions are derived. (The details of computations
can be found in Appendixes A and B). The paper ends with our conclusion in
Section IV.

\section{Formalism for TSWs}

We start with a $3+1-$dimensional flat spacetime in spherical coordinates 
\begin{equation}
ds^{2}=-dt^{2}+dr^{2}+r^{2}\left( d\theta ^{2}+\sin ^{2}\theta d\varphi
^{2}\right) ,
\end{equation}%
and introduce a closed hypersurface defined by%
\begin{equation}
\Sigma \left( t,r,\theta ,\varphi \right) =r-R\left( t,\theta ,\varphi
\right) =0
\end{equation}%
such that the original spacetime is divided into two parts which will form
the inside and outside of the given hypersurface in (2). Now, we get two
copies of the outside manifold and glue them at the hypersurface $\Sigma
\left( t,r,\theta ,\varphi \right) =0.$ What is constructed by this
procedure is a complete manifold and as had been introduced first by M.
Visser it is called a thin-shell wormhole (TSW) whose throat turns out to be
the closed hypersurface $r=R\left( t,\theta ,\varphi \right) $ \cite{MV}.
Let's choose $x^{\alpha }=\left( t,r,\theta ,\varphi \right) $ for the $3+1-$%
dimensional spacetime and $\xi ^{i}=\left( t,\theta ,\varphi \right) $ for
the $2+1-$dimensional hypersurface $\Sigma \left( t,r,\theta ,\varphi
\right) =0$. The induced metric tensor of the hypersurface $h_{ij}$ is
defined by 
\begin{equation}
h_{ij}=\frac{\partial x^{\alpha }}{\partial \xi ^{i}}\frac{\partial x^{\beta
}}{\partial \xi ^{j}}g_{\alpha \beta },
\end{equation}%
which yields%
\begin{multline}
ds_{\Sigma }^{2}=-\left( 1-R_{,t}^{2}\right) dt^{2}+\left( R^{2}+R_{,\theta
}^{2}\right) d\theta ^{2}+ \\
\left( R^{2}\sin ^{2}\theta +R_{,\varphi }^{2}\right) d\varphi
^{2}+2R_{,t}R_{,\theta }dtd\theta \\
+2R_{,t}R_{,\varphi }dtd\varphi +2R_{,\theta }R_{,\varphi }d\theta d\varphi .
\end{multline}%
Note that our notation $R_{,i}$ means partial derivative with respect to $%
x^{i}.$ Next, we apply the Einstein equations on the shell, also called the
Israel junction conditions \cite{Israel} which are%
\begin{equation}
k_{i}^{j}-k\delta _{i}^{j}=-8\pi S_{i}^{j},
\end{equation}%
where $k_{i}^{j}=K_{i}^{j+}-K_{i}^{j-},$ $k=trace\left( k_{i}^{j}\right) ,$ $%
K_{i}^{j\pm }$ are the extrinsic curvatures of the hypersurface in either
sides. $S_{i}^{j}$ is the energy momentum tensor on the shell with the
components 
\begin{equation}
S_{i}^{j}=\left( 
\begin{array}{ccc}
-\sigma & S_{t}^{\theta } & S_{t}^{\varphi } \\ 
S_{\theta }^{t} & p_{\theta }^{\theta } & S_{\theta }^{\varphi } \\ 
S_{t}^{\varphi } & S_{\varphi }^{\theta } & p_{\varphi }^{\varphi }%
\end{array}%
\right)
\end{equation}%
where $\sigma $ is the energy density on the surface and $S_{i}^{j}$ are the
appropriate energy-momentum flux and momentum densities, respectively. Our
explicit calculations reveal (see Appendix A) 
\begin{equation}
k_{tt}=-\frac{2}{\sqrt{\Delta }}R_{,t,t}
\end{equation}%
\begin{equation}
k_{\theta \theta }=-\frac{2}{\sqrt{\Delta }}\left( R_{,\theta ,\theta }-R-%
\frac{2R_{,\theta }^{2}}{R}\right)
\end{equation}%
\begin{multline}
k_{\varphi \varphi }=-\frac{2}{\sqrt{\Delta }}\times \\
\left( R_{,\varphi ,\varphi }-R\sin ^{2}\theta +R_{,\theta }\sin \theta \cos
\theta -\frac{2R_{,\varphi }^{2}}{R}\right)
\end{multline}%
\begin{equation}
k_{t\theta }=-\frac{2}{\sqrt{\Delta }}\left( R_{,t,\theta }-\frac{R_{,\theta
}R_{,t}}{R}\right)
\end{equation}%
\begin{equation}
k_{t\varphi }=-\frac{2}{\sqrt{\Delta }}\left( R_{,t,\varphi }-\frac{%
R_{,\varphi }R_{,t}}{R}\right)
\end{equation}%
and%
\begin{equation}
k_{\theta \varphi }=-\frac{2}{\sqrt{\Delta }}\left( R_{,\theta ,\varphi }-%
\frac{2R_{,\theta }R_{,\varphi }}{R}-R_{,\varphi }\cot \theta \right) .
\end{equation}%
In Appendix B we give the mixed tensor $k_{i}^{j}$ in closed forms which are
also used to determine explicit expressions for the energy momentum tensor.

\section{ Static TSWs}

Using the general formalism given above and in the Appendixes A and B we may
consider some specific cases. First of all we consider the case in which the
throat is static. This means $R\left( t,\theta ,\varphi \right) =\mathcal{R}%
\left( \theta ,\varphi \right) $ and consequently the line element on the
shell / TSW becomes%
\begin{multline}
ds_{\Sigma }^{2}=-dt^{2}+\left( \mathcal{R}^{2}+\mathcal{R}_{,\theta
}^{2}\right) d\theta ^{2}+ \\
\left( \mathcal{R}^{2}\sin ^{2}\theta +\mathcal{R}_{,\varphi }^{2}\right)
d\varphi ^{2}+2\mathcal{R}_{,\theta }\mathcal{R}_{,\varphi }d\theta d\varphi
.
\end{multline}%
The non-zero components of the effective extrinsic curvature tensor are then
given as%
\begin{equation}
k_{\theta \theta }=-\frac{2}{\sqrt{\Delta }}\left( \mathcal{R}_{,\theta
,\theta }-\mathcal{R}-\frac{2\mathcal{R}_{,\theta }^{2}}{\mathcal{R}}\right)
,
\end{equation}%
\begin{multline}
k_{\varphi \varphi }=-\frac{2}{\sqrt{\Delta }}\times \\
\left( \mathcal{R}_{,\varphi ,\varphi }-\mathcal{R}\sin ^{2}\theta +\mathcal{%
R}_{,\theta }\sin \theta \cos \theta -\frac{2\mathcal{R}_{,\varphi }^{2}}{%
\mathcal{R}}\right) ,
\end{multline}%
and%
\begin{equation}
k_{\theta \varphi }=-\frac{2}{\sqrt{\Delta }}\left( \mathcal{R}_{,\theta
,\varphi }-\frac{2\mathcal{R}_{,\theta }\mathcal{R}_{,\varphi }}{\mathcal{R}}%
-\mathcal{R}_{,\varphi }\cot \theta \right) .
\end{equation}%
Finally in static configuration, one finds%
\begin{multline}
-8\pi \sigma _{0}=\frac{2\left( \mathcal{R}^{2}\sin ^{2}\theta +\mathcal{R}%
_{,\varphi }^{2}\right) \left( \mathcal{R}_{,\theta ,\theta }-\mathcal{R}-%
\frac{2\mathcal{R}_{,\theta }^{2}}{\mathcal{R}}\right) }{h\sqrt{\Delta }} \\
-\frac{4\mathcal{R}_{,\theta }\mathcal{R}_{,\varphi }\left( \mathcal{R}%
_{,\theta ,\varphi }-\frac{2\mathcal{R}_{,\theta }\mathcal{R}_{,\varphi }}{%
\mathcal{R}}-\mathcal{R}_{,\varphi }\cot \theta \right) }{h\sqrt{\Delta }} \\
+\frac{2\left( \mathcal{R}^{2}+\mathcal{R}_{,\theta }^{2}\right) }{h\sqrt{%
\Delta }}\times \\
\left( \mathcal{R}_{,\varphi ,\varphi }-\mathcal{R}\sin ^{2}\theta +\mathcal{%
R}_{,\theta }\sin \theta \cos \theta -\frac{2\mathcal{R}_{,\varphi }^{2}}{%
\mathcal{R}}\right) ,
\end{multline}%
\begin{multline}
8\pi p_{\theta }^{\theta }=k_{t}^{t}+k_{\varphi }^{\varphi }=\frac{2\left( 
\mathcal{R}^{2}+\mathcal{R}_{,\theta }^{2}\right) }{h\sqrt{\Delta }}\times \\
\left( \mathcal{R}_{,\varphi ,\varphi }-\mathcal{R}\sin ^{2}\theta +\mathcal{%
R}_{,\theta }\sin \theta \cos \theta -\frac{2\mathcal{R}_{,\varphi }^{2}}{%
\mathcal{R}}\right) \\
-\frac{2\mathcal{R}_{,\theta }\mathcal{R}_{,\varphi }}{h\sqrt{\Delta }}%
\left( \mathcal{R}_{,\theta ,\varphi }-\frac{2\mathcal{R}_{,\theta }\mathcal{%
R}_{,\varphi }}{\mathcal{R}}-\mathcal{R}_{,\varphi }\cot \theta \right) ,
\end{multline}%
and%
\begin{multline}
8\pi p_{\varphi }^{\varphi }=k_{t}^{t}+k_{\theta }^{\theta }= \\
+\frac{2\left( \mathcal{R}^{2}\sin ^{2}\theta +\mathcal{R}_{,\varphi
}^{2}\right) }{h\sqrt{\Delta }}\left( \mathcal{R}_{,\theta ,\theta }-%
\mathcal{R}-\frac{2\mathcal{R}_{,\theta }^{2}}{\mathcal{R}}\right) \\
-\frac{2\mathcal{R}_{,\theta }\mathcal{R}_{,\varphi }}{h\sqrt{\Delta }}%
\left( \mathcal{R}_{,\theta ,\varphi }-\frac{2\mathcal{R}_{,\theta }\mathcal{%
R}_{,\varphi }}{\mathcal{R}}-\mathcal{R}_{,\varphi }\cot \theta \right) .
\end{multline}%
We note that for the static TSW we have%
\begin{equation}
\Delta =1+\frac{\mathcal{R}_{,\theta }^{2}}{\mathcal{R}^{2}}+\frac{\mathcal{R%
}_{,\varphi }^{2}}{\mathcal{R}^{2}\sin ^{2}\theta }
\end{equation}%
and%
\begin{equation}
h=-\mathcal{R}^{2}\left( \sin ^{2}\theta \left( \mathcal{R}^{2}+\mathcal{R}%
_{,\theta }^{2}\right) +\mathcal{R}_{,\varphi }^{2}\right) .
\end{equation}%
Having the exact form of $\sigma _{0}$, we find the total energy which
supports the static TSW. This can be done by using the following integral%
\begin{equation}
\Omega =\int_{0}^{2\pi }\int_{0}^{\pi }\int_{0}^{\infty }\sqrt{-g}\sigma
\delta \left( r-\mathcal{R}\right) drd\theta d\varphi
\end{equation}%
which becomes%
\begin{equation}
\Omega =\int_{0}^{2\pi }\int_{0}^{\pi }\sigma _{0}\mathcal{R}^{2}\sin \theta
d\theta d\varphi
\end{equation}%
upon using the property of the Dirac delta function $\delta \left( r-%
\mathcal{R}\right) $. To make this energy positive we have to consider an
appropriate function for $\mathcal{R}\left( \theta ,\varphi \right) $ and
calculate the total energy $\Omega .$ Particular cases of $\mathcal{R}\left(
\theta ,\varphi \right) $ are considered in the sequel.

\subsection{ $\mathcal{R}\left( \protect\theta ,\protect\varphi \right) $
function of $\protect\theta $ only}

Let's make the simpler choice by considering $\mathcal{R}\left( \theta
,\varphi \right) =\Theta \left( \theta \right) .$ Following this one gets 
\begin{equation}
ds_{\Sigma }^{2}=-dt^{2}+\left( \Theta ^{2}+\Theta _{,\theta }^{2}\right)
d\theta ^{2}+\Theta ^{2}\sin ^{2}\theta d\varphi ^{2},
\end{equation}%
where the only non-zero components of the extrinsic curvature are%
\begin{equation}
k_{\theta \theta }=-\frac{2}{\sqrt{\Delta }}\left( \Theta _{,\theta ,\theta
}-\Theta -\frac{2\Theta _{,\theta }^{2}}{\Theta }\right)
\end{equation}%
and%
\begin{equation}
k_{\varphi \varphi }=-\frac{2}{\sqrt{\Delta }}\left( -\Theta \sin ^{2}\theta
+\Theta _{,\theta }\sin \theta \cos \theta \right) .
\end{equation}%
Consequently,%
\begin{equation}
\tilde{\sigma}_{0}=8\pi \sigma _{0}=\frac{2\left( \frac{\Theta _{,\theta
,\theta }-\Theta -\frac{2\Theta _{,\theta }^{2}}{\Theta }}{\Theta
^{2}+\Theta _{,\theta }^{2}}+\frac{\Theta _{,\theta }\cot \theta -\Theta }{%
\Theta ^{2}}\right) }{\sqrt{1+\frac{\Theta _{,\theta }^{2}}{\Theta ^{2}}}},
\end{equation}%
\begin{equation}
\tilde{p}_{0}=8\pi p_{\theta }^{\theta }=k_{t}^{t}+k_{\varphi }^{\varphi }=%
\frac{2\left( \Theta -\Theta _{,\theta }\cot \theta \right) }{\Theta ^{2}%
\sqrt{1+\frac{\Theta _{,\theta }^{2}}{\Theta ^{2}}}},
\end{equation}%
and%
\begin{equation}
\tilde{q}_{0}=8\pi p_{\varphi }^{\varphi }=k_{t}^{t}+k_{\theta }^{\theta }=-%
\frac{2\left( \Theta _{,\theta ,\theta }-\Theta -\frac{2\Theta _{,\theta
}^{2}}{\Theta }\right) }{\left( \Theta ^{2}+\Theta _{,\theta }^{2}\right) 
\sqrt{1+\frac{\Theta _{,\theta }^{2}}{\Theta ^{2}}}}.
\end{equation}%
One must note that $r=\Theta \left( \theta \right) $ is the hypersurface of
the throat, therefore $\Theta \left( \theta \right) $ must be chosen such
that the surface remains closed. For instance, if we set $\Theta \left(
\theta \right) =a=const.$ then the throat will be a spherical shell of
radius $a$ and $8\pi \sigma _{0}=-\frac{4}{a}$ which is clearly negative and
so is $\Omega .$ Picking more complicated functions periodic in $\theta $ is
acceptable provided it makes the total energy positive. Here, having $\sigma
_{0}\geq 0$ is a sufficient condition to have $\Omega \geq 0,$ but not
necessary. Our main purpose as we stated in the Introduction is to show that
there is possibility of having a TSW supported by ordinary matter in the
sense that $\tilde{\sigma}_{0}\geq 0.$ This condition effectively reduces to%
\begin{equation}
\frac{\Theta _{,\theta ,\theta }-\Theta -\frac{2\Theta _{,\theta }^{2}}{%
\Theta }}{\Theta ^{2}+\Theta _{,\theta }^{2}}+\frac{\Theta _{,\theta }\cot
\theta -\Theta }{\Theta ^{2}}\geq 0.
\end{equation}%
At this stage we shall not purse for a $\Theta \left( \theta \right) $ that
satisfies this condition.

\subsection{$\mathcal{R}\left( \protect\theta ,\protect\varphi \right) $
function of $\protect\varphi $ only}

As in the previous section, here we consider $\mathcal{R}\left( \theta
,\varphi \right) =\Phi \left( \varphi \right) $ which yields%
\begin{equation}
8\pi \sigma _{0}=\frac{2\left( \Phi _{,\varphi ,\varphi }-2\Phi \sin
^{2}\theta -\frac{3\Phi _{,\varphi }^{2}}{\Phi }\right) }{\left( \Phi
^{2}\sin ^{2}\theta +\Phi _{,\varphi }^{2}\right) \sqrt{1+\frac{\Phi
_{,\varphi }^{2}}{\Phi ^{2}\sin ^{2}\theta }}}.
\end{equation}%
If we set $\Phi =a,$ then $8\pi \sigma _{0}=\frac{-4}{a}$ once more as it
should be. It is observed that even with $\mathcal{R}\left( \theta ,\varphi
\right) =\Phi \left( \varphi \right) $ the energy density $\sigma _{0}$ is a
function of both $\theta $ and $\varphi .$ In Fig. 1 we plot 
\begin{equation}
\Phi =\frac{1}{\sqrt{\left\vert \cos \left( 3\varphi \right) \right\vert }+1}
\end{equation}%
which implies that $\sigma _{0}>0$ everywhere. This is what we were looking
for, at least in this stage. We also note that the total energy through a
numerical computation (given by Eq. (23)) is finite, and more importantly,
positive i.e., $\Omega =22.137.$ This shows that the turning / critical
points on the throat do not demand infinite energy and therefore the model
can be physically acceptable. This situation is similar to the case of
2+1-dimensional studied in \cite{MH3}. 
\begin{figure}[h]
\includegraphics[width=80mm,scale=0.7]{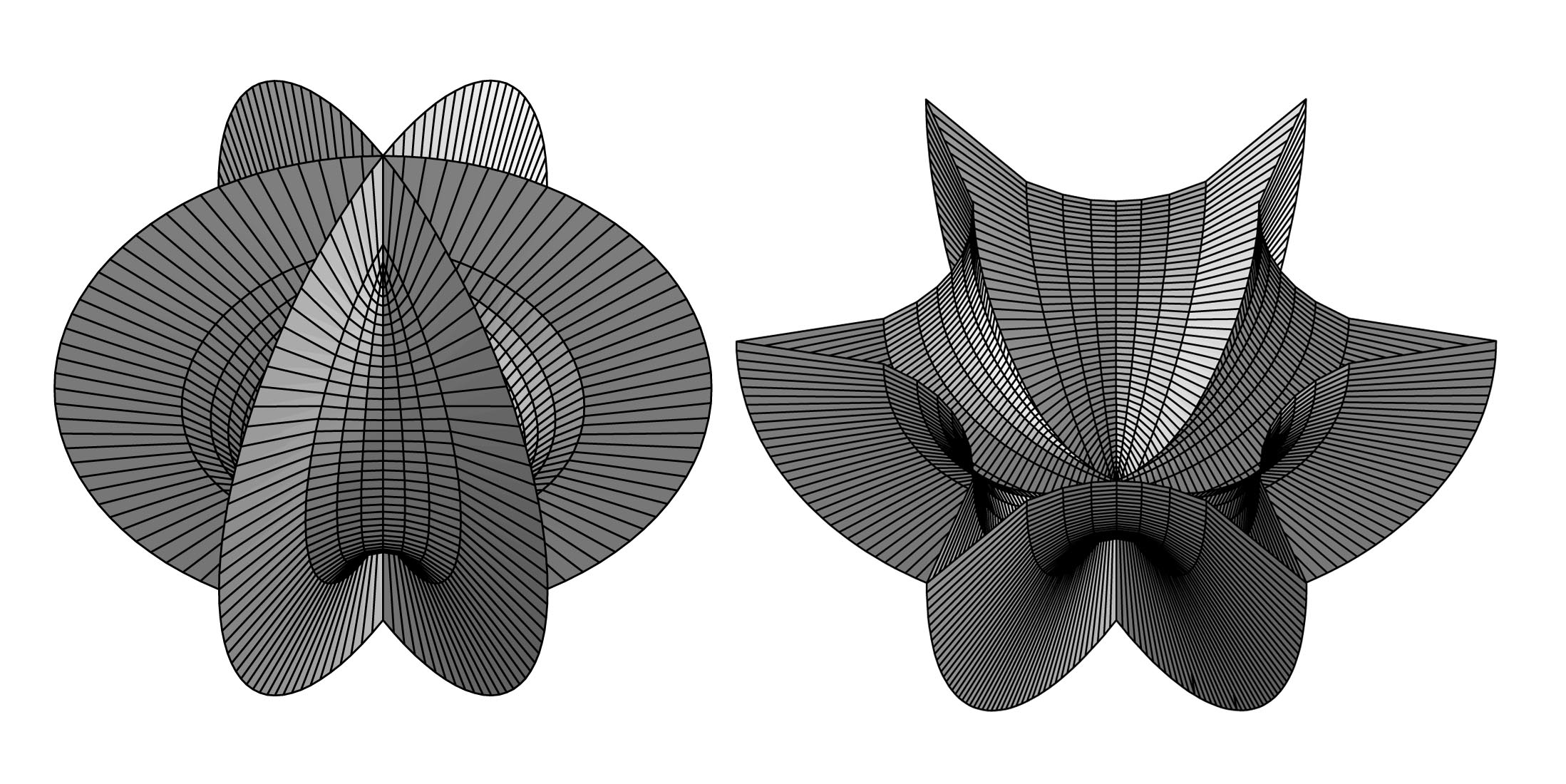}
\caption{A plot (left) of a thin-shell with the hyperplane equation $\Phi =%
\frac{1}{\protect\sqrt{\left\vert \cos \left( 3\protect\varphi \right)
\right\vert }+1}$ which admits the energy density on the shell positive
everywhere. The right figure is the opening of the left which shows that the
surface is concave out everywhere. It should be added that the sharp edges
can be smoothed at the expense of adding negative energy. Since we refrain
doing this we have to face differentiability problem at those edges. We note
that the total energy is finite in these sharp edges. }
\end{figure}


\subsection{$\mathcal{R}\left( \protect\theta ,\protect\varphi \right) $ as
a general periodic function}

Now, we state the most general condition which is provided by a general
periodic function for $\mathcal{R}\left( \theta ,\varphi \right) .$ As a
matter of fact, in (17) we gave in closed form such a $\sigma _{0}$ and what
is left is to provide a proper function for $\mathcal{R}\left( \theta
,\varphi \right) $ such that $\sigma _{0}\geq 0.$ Fig. 2 is a typical
example which admits the energy density $\sigma _{0}$ positive. In this
figure we set%
\begin{equation}
\mathcal{R}\left( \theta ,\varphi \right) =\frac{1}{\left( \sqrt{\left\vert
\cos \left( \theta \right) \right\vert }+1\right) \left( \sqrt{\left\vert
\cos \left( 4\varphi \right) \right\vert }+1\right) },
\end{equation}%
which is dependent on both of the spherical angles. As in Fig. 1, our
numerical calculation reveals that the total energy calculated by Eq. (23)
is finite and positive i.e., $\Omega =28.900.$ This shows that the edges of
the throat are made of a finite amount of energy which is desired in a
physical model.

%
\begin{figure}[h]
\includegraphics[width=80mm,scale=0.7]{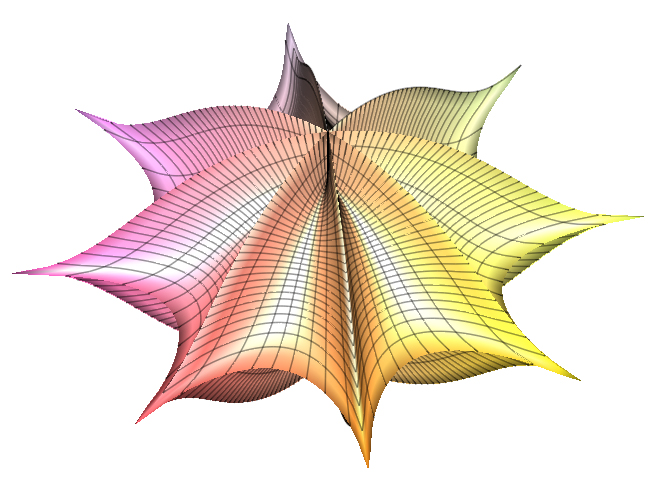}
\caption{A plot of a thin-shell with the equation $\mathcal{R}\left( \protect%
\theta ,\protect\varphi \right) =\frac{1}{\left( \protect\sqrt{\left\vert
\cos \left( \protect\theta \right) \right\vert }+1\right) \left( \protect%
\sqrt{\left\vert \cos \left( 4\protect\varphi \right) \right\vert }+1\right) 
}$ in spherical coordinate system. This specific form of throat provides the
energy density positive on the throat. Similar problems raised about sharp
edges in caption of Fig. 1 are valid also here. The total energy which
supports the throat is finite and positive even at the sharp edges energy
does not diverge.}
\end{figure}


\subsection{Existence of solution}

In \cite{Flareout} we have shown that for the TSW in $3+1-$dimensions, $%
\sigma $ relates to the trace of extrinsic curvature of spatial part of the
Gaussian line element which amounts to 
\begin{equation}
-8\pi \sigma =k_{\theta }^{\theta }+k_{\varphi }^{\varphi }.
\end{equation}%
Therefore for $\sigma \geq 0$ the spatial extrinsic curvature must be
negative, that is why for a positive curvature shape such as a sphere $%
\sigma $ is negative. We note that an open surface with negative curvature
can not be also an answer to our demand, because the throat of a TSW is
defined to be closed. To show that such shapes i.e., negatively curved but
closed, exist, we refer to, for instance, a concave dodecahedron. This is
defined as a surface whose faces are concave individually, like the cellular
surface of a soccer ball with inside pressure less than outside. In such
shapes, although the surface is closed, it consists of negatively curved
individual patches in geometry with anti-de Sitter spacetime and hence makes 
$\sigma \geq 0.$ Similar argument is also valid in $2+1-$dimensions which we
have considered in \cite{MH3}.

\section{Conclusion}

The throat geometry for TSWs is taken embedded in $3+1-$dimensional flat
geometry in spherical coordinates. For static case we obtain the most
general angular dependent constraints the functions have to satisfy in order
to yield a positive total energy. We must admit that the positivity
condition refers to a static frame in which the energy density becomes
positive although the NEC remains violated. Specific reduction procedures
are given dependent on both $\varphi ,$ and $\theta $ and $\varphi $, that
simplify the constraint conditions. Once these constraint conditions are
satisfied we shall not be destined to confront exotic matter in TSWs. At
least in particular, static frames two particular examples are given which
yield positive total energy $\Omega $, from Eq. (23). We admit also that
finding analytically general integrals for functions to satisfy our
differential equation constraints doesn't seem an easy task at all. The
details of our technical part are given in Appendix. The argument / method
can naturally be extended to cover more general wormholes, not only the
TSWs. One issue that remains open, in all this endeavor which we have not
discussed, is the stability of such constructions. A final warning to the
traveller who intends to cross the throat: The thin edges may give harmful
tidal effects from geometrical point of view, so keep away from those edges
if you dream to enjoy a journey at all.

\appendix

\section{Extrinsic Curvature tensor}

To find the induced extrinsic curvature tensor on $\Sigma $ we find the unit
four-normal vector defined as%
\begin{equation}
n_{\gamma }^{\pm }=\pm \frac{1}{\sqrt{\Delta }}\frac{\partial \Sigma \left(
t,r,\theta ,\varphi \right) }{\partial x^{\gamma }}
\end{equation}%
in which 
\begin{equation}
\Delta =\frac{\partial \Sigma \left( t,r,\theta ,\varphi \right) }{\partial
x^{\alpha }}\frac{\partial \Sigma \left( t,r,\theta ,\varphi \right) }{%
\partial x^{\beta }}g^{\alpha \beta }
\end{equation}%
and $\pm $ reefers to the inward and outward directions on the sides of the
hypersurface. An explicit calculation yields%
\begin{equation}
n_{t}^{\pm }=\pm \frac{1}{\sqrt{\Delta }}\left( -R_{,t}\right) ,
\end{equation}%
\begin{equation}
n_{r}^{\pm }=\pm \frac{1}{\sqrt{\Delta }},
\end{equation}%
\begin{equation}
n_{\theta }^{\pm }=\pm \frac{1}{\sqrt{\Delta }}\left( -R_{,\theta }\right)
\end{equation}%
and%
\begin{equation}
n_{\varphi }^{\pm }=\pm \frac{1}{\sqrt{\Delta }}\left( -R_{,\varphi }\right)
.
\end{equation}%
We also find 
\begin{equation}
\Delta =1-R_{,t}^{2}+\frac{R_{,\theta }^{2}}{R^{2}}+\frac{R_{,\varphi }^{2}}{%
R^{2}\sin ^{2}\theta }.
\end{equation}%
The definition of the extrinsic curvature tensor is given by 
\begin{equation}
K_{ij}^{\pm }=-n_{\gamma }^{\pm }\left( \frac{\partial ^{2}x^{\gamma }}{%
\partial \xi ^{i}\partial \xi ^{j}}+\Gamma _{\alpha \beta }^{\gamma }\frac{%
\partial x^{\alpha }}{\partial \xi ^{i}}\frac{\partial x^{\beta }}{\partial
\xi ^{j}}\right) ,
\end{equation}%
in which we have $\Gamma _{r\theta }^{\theta }=\Gamma _{r\varphi }^{\varphi
}=\frac{1}{r},$ $\Gamma _{\theta \theta }^{r}=-r,$ $\Gamma _{\theta \varphi
}^{\varphi }=\cot \theta ,$ $\Gamma _{\varphi \varphi }^{r}=-r\sin
^{2}\theta $ and $\Gamma _{\varphi \varphi }^{\theta }=\sin \theta \cos
\theta .$ One finds%
\begin{equation}
K_{t\theta }^{\pm }=-n_{r}^{\pm }R_{,t,\theta }-n_{\theta }^{\pm }\left( 
\frac{R_{,t}}{R}\right) ,
\end{equation}%
\begin{equation}
K_{t\varphi }^{\pm }=-n_{r}^{\pm }R_{,t,\varphi }-n_{\varphi }^{\pm }\left( 
\frac{R_{,t}}{R}\right) ,
\end{equation}%
\begin{equation}
K_{\theta \varphi }^{\pm }=-n_{r}^{\pm }R_{,\theta ,\varphi }-n_{\theta
}^{\pm }\left( \frac{R_{,\varphi }}{R}\right) -n_{\varphi }^{\pm }\left( 
\frac{R_{,\theta }}{R}+\cot \theta \right) ,
\end{equation}%
\begin{equation}
K_{tt}^{\pm }=-n_{r}^{\pm }R_{,t,t},
\end{equation}%
\begin{equation}
K_{\theta \theta }^{\pm }=-n_{r}^{\pm }\left( R_{,\theta ,\theta }-R\right)
-2n_{\theta }^{\pm }\frac{R_{,\theta }}{R}
\end{equation}%
and%
\begin{equation}
K_{\varphi \varphi }^{\pm }=-n_{r}^{\pm }\left( R_{,\varphi ,\varphi }-R\sin
^{2}\theta \right) +n_{\theta }^{\pm }\sin \theta \cos \theta -2n_{\varphi
}^{\pm }\frac{R_{,\varphi }}{R}.
\end{equation}

\section{Energy momentum tensor}

We start with%
\begin{equation}
k_{t}^{t}=h^{tt}k_{tt}+h^{t\theta }k_{t\theta }+h^{t\varphi }k_{t\varphi }
\end{equation}%
\begin{equation}
k_{\theta }^{\theta }=h^{\theta \theta }k_{\theta \theta }+h^{\theta
t}k_{\theta t}+h^{\theta \varphi }k_{\theta \varphi }
\end{equation}%
\begin{equation}
k_{\varphi }^{\varphi }=h^{\varphi \varphi }k_{\varphi \varphi }+h^{\varphi
t}k_{\varphi t}+h^{\varphi \theta }k_{\varphi \theta }
\end{equation}%
\begin{equation}
k_{t}^{\theta }=h^{\theta t}k_{tt}+h^{\theta \theta }k_{t\theta }+h^{\theta
\varphi }k_{t\varphi }
\end{equation}%
\begin{equation}
k_{\theta }^{t}=h^{tt}k_{\theta t}+h^{t\theta }k_{\theta \theta
}+h^{t\varphi }k_{\theta \varphi }
\end{equation}%
\begin{equation}
k_{t}^{\varphi }=h^{\varphi t}k_{tt}+h^{\varphi \theta }k_{t\theta
}+h^{\varphi \varphi }k_{t\varphi }
\end{equation}%
\begin{equation}
k_{\varphi }^{t}=h^{tt}k_{\varphi t}+h^{t\theta }k_{\varphi \theta
}+h^{t\varphi }k_{\varphi \varphi }
\end{equation}%
\begin{equation}
k_{\theta }^{\varphi }=h^{\varphi \theta }k_{\theta \theta }+h^{\varphi
t}k_{\theta t}+h^{\varphi \varphi }k_{\theta \varphi }
\end{equation}%
and%
\begin{equation}
k_{\varphi }^{\theta }=h^{\theta \varphi }k_{\varphi \varphi }+h^{\theta
t}k_{\varphi t}+h^{\theta \theta }k_{\varphi \theta }.
\end{equation}%
Let's introduce the metric tensor%
\begin{equation}
\mathbf{h}=\left( 
\begin{array}{ccc}
-\left( 1-R_{,t}^{2}\right) & R_{,t}R_{,\theta } & R_{,t}R_{,\varphi } \\ 
R_{,t}R_{,\theta } & R^{2}+R_{,\theta }^{2} & R_{,\theta }R_{,\varphi } \\ 
R_{,t}R_{,\varphi } & R_{,\theta }R_{,\varphi } & R^{2}S^{2}+R_{,\varphi
}^{2}%
\end{array}%
\right)
\end{equation}%
and its inverse%
\begin{multline}
\mathbf{h}^{-1}= \\
\left( 
\begin{array}{ccc}
\frac{\left[ \left( R^{2}+R_{,\theta }^{2}\right) S^{2}+R_{,\varphi }^{2}%
\right] R^{2}}{h} & \frac{R^{2}R_{,t}R_{,\theta }S^{2}}{h} & \frac{%
R^{2}R_{,t}R_{,\varphi }}{h} \\ 
\frac{R^{2}R_{,t}R_{,\theta }S^{2}}{h} & \frac{R^{2}\left(
R_{,t}^{2}-1\right) S^{2}-R_{,\varphi }^{2}}{h} & \frac{R_{,\theta
}R_{,\varphi }}{h} \\ 
\frac{R^{2}R_{,t}R_{,\varphi }}{h} & \frac{R_{,\theta }R_{,\varphi }}{h} & 
\frac{R^{2}\left( R_{,t}^{2}-1\right) -R_{,\theta }^{2}}{h}%
\end{array}%
\right)
\end{multline}%
in which $S=\sin \theta $ and $h$ is the determinant of $\mathbf{h}$ i.e., 
\begin{equation}
h=-R^{2}\left( \sin ^{2}\theta \left( R^{2}\left[ 1-R_{,t}^{2}\right]
+R_{,\theta }^{2}\right) +R_{,\varphi }^{2}\right) .
\end{equation}%
Considering the Israel junction conditions we find%
\begin{multline}
-8\pi \sigma =k_{\theta }^{\theta }+k_{\varphi }^{\varphi }= \\
-\frac{2\left( R^{2}\left( R_{,t}^{2}-1\right) \sin ^{2}\theta -R_{,\varphi
}^{2}\right) \left( R_{,\theta ,\theta }-R-\frac{2R_{,\theta }^{2}}{R}%
\right) }{h\sqrt{\Delta }}- \\
\frac{2R^{2}R_{,t}R_{,\theta }\sin ^{2}\theta \left( R_{,t,\theta }-\frac{%
R_{,\theta }R_{,t}}{R}\right) }{h\sqrt{\Delta }}- \\
\frac{4R_{,\theta }R_{,\varphi }\left( R_{,\theta ,\varphi }-\frac{%
2R_{,\theta }R_{,\varphi }}{R}-R_{,\varphi }\cot \theta \right) }{h\sqrt{%
\Delta }} \\
-\frac{2\left( R^{2}\left( R_{,t}^{2}-1\right) -R_{,\theta }^{2}\right) }{h%
\sqrt{\Delta }}\times \\
\left( R_{,\varphi ,\varphi }-R\sin ^{2}\theta +R_{,\theta }\sin \theta \cos
\theta -\frac{2R_{,\varphi }^{2}}{R}\right) - \\
\frac{2R^{2}R_{,t}R_{,\varphi }\left( R_{,t,\varphi }-\frac{R_{,\varphi
}R_{,t}}{R}\right) }{h\sqrt{\Delta }}.
\end{multline}%
The other components of the energy momentum tensor can be found similarly.

\bigskip


\begin{thebibliography}{99}
\bibitem{MT} M. S. Morris and K. S. Thorne, Am. J. Phys. \textbf{56}, 395
(1988).

\bibitem{V} M. Visser, Phys. Rev. D \textbf{39}, 3182 (1989).

\bibitem{Rotating} K. A. Bronnikov, V. G. Krechet and J. P.S. Lemos, Phys.
Rev. D \textbf{87}, 084060 (2013);

K.A. Bronnikov and J. P. S. Lemos, Phys. Rev. D \textbf{79}, 104019 (2009);

\bibitem{Stability} P. K. F. Kuhfittig, Fundamental J. Mod. Phys. \textbf{7}%
, 111 (2014);

C. Bejarano, E. F. Eiroa and C. Simeone, Eur. Phys. J. C \textbf{74}, 3015
(2014);

S. H. Mazharimousavi, M. Halilsoy and Z. Amirabi, Phy. Rev. D \textbf{89},
084003 (2014);

A. Banerjee, Int. J. Theor. Phys. 52, 2943 (2013);

M. Sharif and M. Azam, JCAP 05, 25 (2013);

M. Sharif and M. Azam, Eur. Phys. J. C \textbf{73}, 2407 (2013);

M. Sharif and M. Azam, J. Phys. Soc. Jpn. \textbf{81}, 124006 (2012);

M. Sharif and M. Azam, JCAP \textbf{04}, 23 (2013);

M. H. Dehghani and M. R. Mehdizadeh, Phys. Rev. D \textbf{85}, 024024 (2012);

X. Yue and S. Gao, Phys. Lett. A \textbf{375}, 2193 (2011);

P. K. F. Kuhfittig, Acta Phys. Polon. B \textbf{41, }2017 (2010);

J. P. S. Lemos and F. S. N. Lobo, Phys. Rev. D \textbf{78}, 044030 (2008);

E. F. Eiroa, Phys. Rev. D \textbf{78, }024018 (2008);

E. F. Eiroa and C. Simeone, Phys. Rev. D \textbf{76}, 024021 (2007);

M. Ishak, K. Lake, Phys. Rev. D\textbf{\ 65,} 044011 (2002);

F. S. N. Lobo and Paulo Crawford, Class. Quant. Grav. \textbf{21,} 391
(2004);

E. Poisson and M. Visser, Phys. Rev. D \textbf{52, }7318 (1995);

\bibitem{Ghost} J. A. Gonzalez, F. S. Guzman and O. Sarbach, Class. Quantum
Grav. \textbf{26}, 015010 (2009);

J. A. Gonzalez, F.S. Guzman and O. Sarbach, Phys. Rev. D \textbf{80}, 024023
(2009).

\bibitem{GB} M. Richarte and C. Simeone, Phys. Rev. D \textbf{76, }087502
(2007); Erratum-ibid.D \textbf{77}, 089903 (2008);

S. H. Mazharimousavi, M. Halilsoy and Z. Amirabi, Phys. Rev. D \textbf{81},
104002 (2010);

S. Habib Mazharimousavi, M. Halilsoy and Z. Amirabi, Class. Quantum Grav. 
\textbf{28}, 025004 (2011);

M. G. Richarte, Phys. Rev. D \textbf{82}, 044021 (2010);

M. G. Richarte, Phys. Rev. D \textbf{87}, 067503 (2013).

\bibitem{ZV} H. Weyl, Ann. Physik, \textbf{54}, 117 (1917);

D.\thinspace M. Zipoy, J. Math. Phys. (N.Y.) \textbf{7}, 1137 (1966);

B. \thinspace H. Voorhees, Phys. Rev. D \textbf{2}, 2119 (1970).

\bibitem{MH} S. H. Mazharimousavi and M. Halilsoy, Eur. Phys. J. C \textbf{74%
}, 3067 (2014);

S. H. Mazharimousavi and M. Halilsoy, Phys. Rev. D \textbf{90}, 087501
(2014).

\bibitem{HV} D. Hochberg and M. Visser, Phys. Rev. D \textbf{56}, 4745
(1997).

\bibitem{MH2} S. H. Mazharimousavi and M. Halilsoy, Eur. Phys. J. C \textbf{%
74}, 3073 (2014).

\bibitem{MH3} S. H. Mazharimousavi and M. Halilsoy, Eur. Phys. J. C \textbf{%
75}, 81 (2015)

\bibitem{MV} M. Visser, Nucl. Phys. \textbf{B} 328, 203 (1989);

P. R. Brady, J. Louko and E. Poisson, Phys. Rev. D \textbf{44}, 1891 (1991);

E. Poisson and M. Visser, Phys. Rev. D \textbf{52}, 7318 (1995);

M. Visser, \textit{Lorentzian Wormholes from Einstein to Hawking} (American
Institute of Physics, New York, 1995).

\bibitem{Israel} W. Israel, Nuovo Cimento \textbf{44B}, 1 (1966);

V. de la Cruz and W. Israel, Nuovo Cimento \textbf{51A}, 774 (1967);

J. E. Chase, Nuovo Cimento \textbf{67B}, 136. (1970);

S. K. Blau, E. I. Guendelman and A. H. Guth, Phys. Rev. D \textbf{35}, 1747
(1987);

R. Balbinot and E. Poisson, Phys. Rev. D \textbf{41}, 395 (1990).

\bibitem{Flareout} S. H. Mazharimousavi and M. Halilsoy, Phys. Rev. D 
\textbf{90}, 087501 (2014).
\end{thebibliography}
\end{document}